\documentclass[12pt]{article}

\usepackage{scicite}
\usepackage{longtable}
\usepackage{times}
\usepackage{epsfig}

\newenvironment{sciabstract}{%
\begin{quote} \bf}
{\end{quote}}

\title{A universal scaling of condensation temperature in quantum fluids}

\author
{S.V. Dordevic,$^{1\ast}$ \\
\\
\normalsize{$^{1}$Department of Physics, The University of Akron,}\\
\normalsize{Akron, OH 44325, USA}\\
\\
\normalsize{$^\ast$To whom correspondence should be addressed; E-mail:  dsasa@uakron.edu}
}

\date{}

\begin{document} 


\baselineskip24pt


\maketitle


\begin{sciabstract}
The phenomena of superconductivity and superfluidity are
believed to originate from the same underlying physics, namely the condensation
of either bosons or pairs of fermions (Cooper pairs). 
In this work I complied and analyzed literature 
data for a number of quantum fluids and showed that indeed they all follow
the same simple scaling law. The critical temperature for condensation T$_c$
is found to scale with the condensate coherence length $\xi$ and 
the effective mass of condensing particles m$^{\ast}$. 
The scaling plot includes members of most known classes
of superconductors, as well as a number of superfluids and condensates, 
such as $^3$He, $^4$He, dilute Bose and Fermi gases, excitons, polaritons, 
neutron superfluid and proton superconductor in neutron stars, nuclear pairing, 
quark--antiquark condensate and Higgs condensate.  The scaling
plot spans more that 24 orders of magnitude of critical temperatures, 
albeit the scaling exponent is not the one predicted by theory.
The plot might help the search for a QCD axion. 
\end{sciabstract}


\section*{The scaling}

The scaling is based on the Ginzburg--Landau theory of second order phase transitions 
\cite{annett-book,leggett-book}. 
In its simplest form, the theory postulates that the Helmholtz free energy density 
of a quantum system undergoing condensation is given as \cite{annett-book,leggett-book}:

\begin{equation}
f_s(T) = f_n(T) + \frac{\hbar^2}{2 m^{\ast}}|\nabla \Psi (\mathbf{r})|^2 + 
\alpha (T)| \Psi(\mathbf{r})|^2 + \frac{\beta (T)}{2} |\Psi(\mathbf{r})|^4 
\label{eq:GL}
\end{equation}
where $\Psi(\mathbf{r})$ is the condensate wave-function, and $f_n$ and $f_s$ 
are the free energy densities in the normal and condensed (superfluid) phase, respectively.
$m^*$ is the mass of condensing particles. For superconductors and fermionic
superfluids $m^{\ast}$ is the mass of a Cooper pair, i.e. 
twice the effective mass of a normal state particle m$_{eff}$: $m^{\ast}$=2m$_{eff}$. 
For bosonic condensates m$^{\ast}$=m$_{eff}$.
Close to transition temperature T$_c$, the phenomenological coefficients $\alpha(T)$ and
$\beta(T)$ are usually approximated as $\alpha(T) \approx a(T - T_c) $ and $\beta (T) \approx  b$ , 
where $a$ and $b$ are temperature-independent constants \cite{a-coef}. 
The second and third terms on the right-hand side of Eq.~\ref{eq:GL} can be interpreted as
kinetic and potential energies of the condensate, respectively. One can 
obtain a simple estimate of the condensation temperature T$_c$ by requiring these 
two terms to be equal: 

\begin{equation}
a T_c = \frac{\hbar^2}{2 m^* \xi^2}
\label{eq:Tc}
\end{equation}
where $\xi$ is the Ginzburg--Landau coherence length of the condensate at T=~0. 
It is also known in the literature as the healing or correlation length, and
it is one of the most fundamental length scales in superconductors and superfluids
\cite{annett-book,leggett-book,tinkham-book}. In superconductors one also defines 
Pippard coherence length, which quantifies the size of a Cooper pair. 
In clean superconductors, well bellow $T_c$, the two coherence
lengths differ only by a numerical factor of order one \cite{tinkham-book}. 
In dirty superconductors the difference between the two can be larger. 

I show below that simple scaling relation between T$_c$, m$^{\ast}$ and $\xi$ 
(Eq.~\ref{eq:Tc}) is followed by all known superconductors, superfluids and 
condensates for which the data is currently available.

\section*{Superconductors}

A scaling between T$_c$, m$^{\ast}$ and $\xi$ for heavy fermion and cuprate 
superconductors was discussed previously  
on phenomenological grounds \cite{angilella00}. I have extended the scaling 
to include members of all known families of superconductors for which the 
data is currently available \cite{hirsch15}. The scaling plot (Fig.~1)
includes: elemental superconductors (such as Pb, Al, Nb, etc.), cuprates 
(both electron doped NCCO and hole doped LSCO, YBCO and Bi-2212, Hg-1201 
and Tl-1222), pnictides, bismuthates, organic superconductors, 
heavy fermions, transition metal dichalcogenides, A15 
superconductors, alkali-doped C$_{60}$ (buckyballs), Chevrel phases, MgB$_2$, 
quaternary borocarbides, topological superconductors, Sr$_2$RuO$_4$, SrTiO$_3$
and others \cite{hirsch15}. With T$_c$ ranging from about 1~K
to about 100~K, they are all located in the lower-right part of the 
plot (red circles in Fig.~1). 

The plot also includes members 
of several superconductor classes that currently attract a lot 
of attention, such as nickelates \cite{nickelate-coh,nickelate-mass} 
and Kagome materials \cite{kagome},  
as well as hydrates (such as H$_2$S \cite{drozdov15}) 
which are superconducting only under very high pressures. 
Although their T$_c$ is low (less than 2~K) magic-angle twisted bilayer 
and trilayer graphene (MATBG and MATTG) \cite{matbg,mattg}, 
as well twisted bilayer WSe$_2$ \cite{wse2},
are important as they shows that quasi-2D superconductors also follow 
the same scaling. Particularly relevant for 
the scaling are the so-called low-T$_c$ superconductors, such as 
Bi (T$_c$= 0.53~mK) and YbRh$_2$Si$_2$ (T$_c$= 7.9~mK), as they extend the 
plot to the right by more than three orders of magnitude.  
Finally, the plot also includes Li$_x$ZrNCl, a quasi--2D system
in which the so-called BCS--BEC crossover \cite{levin24} was observed \cite{ZrNCl}. 

One must be warned that with
such a large collection of data obtained from a variety
of different sources one can expect significant  
scatter of data points. The importance of obtaining consistent sets
of parameters for scaling plots was demonstrated previously 
\cite{Dordevic2013}, where it was shown that parameters obtained on
different samples and with different experimental techniques can 
lead to incorrect conclusions. For the scaling given by
Eq.~\ref{eq:Tc} it is not possible to obtain all three parameters
(T$_c$, m$^*$ and $\xi$) using the same experimental technique \cite{coh-exp}.
Therefore, the values used in Fig.~1 should only be taken as
order of magnitude estimates.

\section*{Superfluids and condensates}

The phenomenon of superfluidity was first discovered in liquid $^4$He at 2.17~K, 
almost a century ago and later explained by Bose-Einstein condensation (BEC). 
On the other hand, in $^3$He fermionic superfluidity was discovered much later 
and at a much lower temperature (2.49~mK). Here the effect was explained 
by p-wave Cooper pairing and BCS-like condensation \cite{leggett-book}.  
The experimental parameters for the scaling 
relation Eq.~\ref{eq:Tc} are readily available for both these quantum
fluids and they are located in the lower-right part of the scaling plot. 
The scaling parameters for $^3$He were also studied in details
as a function of pressure \cite{he3-pressure}.

The last few decades have seen a surge of interests in quantum
fluids, especially after BEC was experimentally realized
in dilute bosonic gasses \cite{bec1,bec2}, followed by condensation of
fermionic gasses \cite{jin03}. 
The critical temperature for bosonic gasses (such as $^7$Li, $^{23}$Na, $^{87}$Rb
and $^1$H) is typically in the $\mu$K range \cite{kleppner98}, 
whereas for fermionic gasses, such as $^6$Li and $^{40}$K, it is in the nK range \cite{fermi-gases}. 
They are all located in the lower-right corner of the scaling plot. 
One should point out that in $^6$Li superfluidity was observed on both sides 
of the BEC--BCS crossover \cite{levin24}, as 
demonstrated by the formation of vortices \cite{bec-bcs},

A number of excitations (quasiparticles) in solids have been predicted 
to undergo condensation, such as excitons, polaritons and 
magnons \cite{universal-themes}. 
There is now a significant body of experimental evidence for the existence
of their condensates.  With a critical temperature of about 1~K, 
exciton condensate \cite{High2012} is located on the lower-right side of 
the plot. On the other hand, polaritons typically have a much smaller 
effective mass \cite{polariton-mass,polariton-length}, which results in 
higher condensation temperatures. Their condensation and superfluidity
have been reported at room temperature \cite{polariton-Tc}, and based 
on the scaling plot I hypothesize that the critical 
temperature could be several times higher. There are other 
quasiparticles in solids which 
are believed to undergo condensation, but for which 
the required parameters are not available at the moment, 
such as magnons, phonons, polarons and plasmons.

Neutron superfluidity was theoretically predicted to exist before neutron stars 
were even discovered \cite{bcs50}. Since then numerous
studies have been performed and values of scaling parameters are known fairly well. 
With transition temperatures on the order of 1~MeV \cite{bcs50}, both neutron superfluid and proton 
superconductor are located in the upper-central part of the plot. 
Other, more exotic condensates, such as pion and quark condensates
(i.e. color superconductivity \cite{color}), might also be present at the cores of 
neutron stars \cite{bcs50}.

Pairing of nucleons inside the nucleus is believed to occur by analogy 
with Cooper pairing and BCS condensation \cite{nuclear_sf}.  
Signatures of the resulting nuclear superfluidity  
were observed experimentally in the form of a pairing gap. 
However, because of the finite size 
effects, the transition to condensed state is smoothed out.
The required scaling parameters are known, and they place nuclear pairing 
on the scaling line.

The so-called QCD phase transition is believed to have occurred in the early Universe
at around 150~MeV (around 10$^{12}$~K). At that temperature chiral symmetry was
broken, resulting in the quark--antiquark condensate \cite{quark-gluon-book}. 
Assuming that neutral pion is the condensing particle, and with the coherence length of about 1~fm, 
one can place the QCD condensate on the scaling line.

Electroweak theory predicts \cite{griffiths-book} a phase transition at around 160~GeV 
(around 10$^{15}$~K) during which the Higgs field acquired a non-zero vacuum
expectation value, i.e. it formed a condensate.  
With the now known mass of the Higgs boson (126~GeV), and the coherence
length of around 10$^{-18}$~m \cite{coleman-book}, 
the Higgs condensate is fully consistent with the scaling. 
It has the highest condensation temperature on the plot,
making the T$_c$ span more than 24 orders of magnitude.

The overall trend in the scaling plot is clear. All known superfluids and 
condensates for which the data is currently available lie on or close to 
the scaling line. 
Surprisingly, however, the power-law behavior is different from the one 
suggested by Eq.~\ref{eq:Tc}. The gray line in Fig.~1 has the slope of 
approximately -0.94$\pm$0.01, instead of -1 expected from Eq.\ref{eq:Tc}. 
Its origin and significance are currently unknown. Naturally, there is 
scattering of data points, as expected from the discussion above.
Several statistical tests were preformed on the scaling parameters
(on $\log(T_c$) versus $\log(m^{\ast}\xi^2$)),
and they all indicate statistically significant correlation: Pearson's 
and Spearman's tests yielded -0.98 and -0.94, respectively.

\section*{Axion} 

A hypothetical particle named axion was proposed almost half a century ago as a 
solution to the strong CP problem \cite{adm-theory}. In recent years
it has also emerged as a leading candidate for a dark matter particle \cite{dm-comm}.
However, numerous experimental searches have failed to find it, in part becuase its
mass is not known. The scaling plot might help us narrow
down the values of its parameters, in particular the mass. The so-called 
QCD axion is believed to have acquired mass at the QCD phase transition 
(crossover) \cite{adm-theory}, at approximately 150~MeV.
If by analogy with the Higgs condensate (see above) one assumes 
that the coherence length of the condensate is given by the axion's reduced Compton 
wavelength \cite{coleman-book,volovik},
one can estimate the effective mass from the plot to be in the range 10--100~MeV \cite{x17}.
(This would correspond to a reduced Compton wavelength of about 20--2~fm.) 
On the other hand, if one uses the effective mass of the order of 1~$\mu$eV 
(the so-called "invisible" axion), as most theoretical
approaches assume \cite{adm-theory} and experimental searches focus on 
\cite{adm-exp,search-ulbdm,sikivie21}, one can estimate from the scaling plot the coherence 
length of about 30~nm. (The reduced Compton wavelength of 
this "invisible" axion would be about 0.2~m.) Other estimates can be made from the plot,
which might help limit the allowed values of parameters 
and hopefully aid the search for a QCD axion.

\section*{Outlook}

A scaling relationship between critical temperature, coherence length
and effective mass was shown to be followed by a variety of
superconductors, superfluids and condensates.  Assuming the 
scaling has a universal character, one can use it to make predictions about 
other systems which have been theorized to undergo condensation, such as 
color superconductivity \cite{color}, anyon superconductivity \cite{anyon}, 
neutrino condensate \cite{neutrino}, photon condensate \cite{photon}, 
phonon condensate \cite{phonons}, gluon condensate \cite{gluons}, 
graviton condensate \cite{universal-themes}, positronium condensate \cite{positronium}, 
supersolidity \cite{supersolid}, and others. 

One might also speculate about the implications the scaling plot might 
have at the Planck scale, which is believed to play an important 
role in quantum gravity \cite{planck-book}.  
Assuming that Planck units ($T_{P}$ = $\sqrt{ \hbar c^5/ G k_B^2}$, 
$m_{P}$ = $\sqrt{\hbar c/ G}$ and $l_{P}$ = $\sqrt{ \hbar G/ c^3} $)   
characterize some unknown phase transition in the early Universe, 
one can easily check that they satisfy Eq.~\ref{eq:Tc}, with a=$\frac{1}{2}k_B$. 
(Planck length is the reduced Compton wavelength 
of Planck mass: $l_P$=$\hbar$/(c$m_P$).) Extending the speculation to GUT scale \cite{gut}, 
one can use the scaling plot to estimate the mass of a hypothetical X boson \cite{gut}
to be about 2$\times$10$^{16}$~GeV and the corresponding coherence length 
(i.e. reduced Compton wavelength) of about 10$^{-32}$~m.


%
\bibliography{mybib}

\begin{thebibliography}{10}
\bibitem{coh-exp}
In superconductors the coherence length can be extracted from several different experimental probes, such as magneto-resistivity, susceptibility and nernst effect measurements, whereas the effective mass can be extracted from specific heat measurements, infrared and microwave spectroscopies, de haas - van alphen measurements, etc.
\bibitem{dm-comm}
There have been several suggestions that dark matter might be in the condensed and/or superfluid phase \cite{silverman01,sikivie09,guth15,khoury15}. however, the parameters (t$_c$, m$^*$ and $\xi$) that some of these models assume or predict fail the scaling by several orders of magnitude, which raises the question of their reliability.
\end{thebibliography}
\bibliographystyle{Science}

\section*{Acknowledgments}

The author thanks P.J. Hirschfeld and D.S.M. Alves for valuable comments on the manuscript. 



\newpage

\section*{Supplementary materials}



\begin{longtable}{|l|c|c|c|l|}
\caption{The critical temperature T$_c$, normal state effective mass 
$m_{eff}$, and coherence length $\xi$ for various superconductors.} \label{tab:scdatabase} \\

\hline \multicolumn{1}{|c|}{\textbf{Superconductor}} & \multicolumn{1}{c|}{\textbf{$T_c$ ($K$)}} & \multicolumn{1}{c|}{\textbf{$m_{eff}$ ($kg$)}} & 
\multicolumn{1}{c|}{\textbf{$\xi$ ($m$)}} & Reference \\ \hline 
\endfirsthead

\multicolumn{5}{c}%
{{\bfseries \tablename\ \thetable{} -- continued from previous page}} \\
\hline \multicolumn{1}{|c|}{\textbf{Superconductor}} & \multicolumn{1}{c|}{\textbf{$T_c$ ($K$)}} & \multicolumn{1}{c|}{\textbf{$m_{eff}$ [$kg$]}} & 
\multicolumn{1}{c|}{\textbf{$\xi$ [$m$]}} & Reference \\ \hline 
\endhead

\hline \multicolumn{5}{|c|}{{Continued on next page}} \\ \hline
\endfoot

\hline \hline
\endlastfoot

 Al         & 1.18    &  1.35$\times$10$^{-30}$    &   1.6$\times$10$^{-6}$    &  \cite{poole-book,kittel-book}        \\ 
 Pb	        & 7.2     &  1.79$\times$10$^{-30}$    &   8.2$\times$10$^{-8}$    &     \\
 Nb	        & 9.25    &  4.43$\times$10$^{-30}$    &   3.9$\times$10$^{-8}$    &     \\ 
 Cd         & 0.56    &  6.64$\times$10$^{-31}$    &   1.1$\times$10$^{-7}$    &      \\ 
 In	        & 3.41    &  1.25$\times$10$^{-30}$    &   3.6$\times$10$^{-7}$    &     \\
 Sn	        & 3.72    &  1.15$\times$10$^{-30}$    &   1.8$\times$10$^{-7}$    &     \\ 
\hline
 NbSe$_2$         & 7.2    &  5.6$\times$10$^{-31}$    &   7.7$\times$10$^{-9}$    &  \cite{buckel-book,poole-book}        \\ 
 MgB$_2$	          & 39     &  4.6$\times$10$^{-31}$    &   5$\times$10$^{-9}$    &     \\
 K$_3$C$_{60}$	        & 19    &  2.2$\times$10$^{-30}$    &   3.4$\times$10$^{-9}$    &     \\ 
 Rb$_3$C$_{60}$         & 30    &  2.1$\times$10$^{-30}$    &   3$\times$10$^{-9}$    &          \\ 
 Nb$_3$Sn	        & 18    &  1.4$\times$10$^{-30}$    &   3$\times$10$^{-9}$    &     \\
 Nb$_3$Ge	        & 23.2    &  1.4$\times$10$^{-30}$    &   3$\times$10$^{-9}$    &     \\ 
 V$_3$Si         & 16    &  1.4$\times$10$^{-30}$    &   3$\times$10$^{-9}$    &          \\ 
 V$_3$Ga	          & 15     &  1.4$\times$10$^{-30}$    &   2.5$\times$10$^{-9}$    &     \\
 YNi$_2$B$_2$C	        & 15.5    &  3.6$\times$10$^{-31}$    &   8$\times$10$^{-9}$    &     \\ 
 LuNi$_2$B$_2$C         & 16.5    &  2.7$\times$10$^{-31}$    &   6$\times$10$^{-9}$    &          \\ 
 TmNi$_2$B$_2$C	        & 11    &  3$\times$10$^{-31}$    &   1.5$\times$10$^{-8}$    &     \\
 Ba$_{0.6}$K$_{0.4}$BiO$_{3}$     & 35    &  3.6$\times$10$^{-30}$    &   3.7$\times$10$^{-9}$    &     \\ 
 PbMo$_6$S$_8$         & 15    &  9.1$\times$10$^{-30}$    &   2.3$\times$10$^{-9}$    &          \\ 
 AgSnSe$_2$	        & 4.55    &  1.37$\times$10$^{-30}$    &   1.25$\times$10$^{-8}$    &     \\
\hline
Sr$_2$RuO$_4$	   & 1.5    &  6.37$\times$10$^{-30}$    &   6.6$\times$10$^{-8}$    &  \cite{sr2ruo4}   \\
SrTiO$_3$	       & 0.5    &  1.66$\times$10$^{-30}$      &    1$\times$10$^{-8}$    &  \cite{sto}   \\
La$_3$Ni$_2$O$_7$  & 62    &     1.82$\times$10$^{-30}$   &   1.84$\times$10$^{-9}$  &  \cite{nickelate-coh,nickelate-mass}  \\ 
CsV$_3$Sb$_5$      & 3.1    &     1.8$\times$10$^{-30}$  &   3.31$\times$10$^{-8}$   &  \cite{kagome}   \\ 	          
\hline
 CeCu$_2$Si$_2$         & 0.7    &  3.46$\times$10$^{-28}$    &   9$\times$10$^{-9}$    &  \cite{angilella00}        \\ 
 URu$_2$Si$_2$	        & 1.2    &  1.27$\times$10$^{-28}$    &   1$\times$10$^{-8}$    &     \\
 UPd$_2$Al$_2$	        & 2      &  6$\times$10$^{-29}$        &   8.5$\times$10$^{-9}$    &     \\ 
 UNi$_2$Al$_2$          & 1    &  4.37$\times$10$^{-29}$        &   2.4$\times$10$^{-8}$    &          \\ 
 UPt$_3$	                & 0.55    &  1.64$\times$10$^{-28}$    &   1$\times$10$^{-8}$    &     \\
 UBe$_{13}$	           & 0.9    &  2.37$\times$10$^{-28}$       &   1$\times$10$^{-8}$    &     \\ 
 CeRh$_2$Si$_2$         & 0.35    &  2$\times$10$^{-28}$    &   3.7$\times$10$^{-8}$    &         \\ 
 CeCoIn$_5$	          & 2.3     &  7.55$\times$10$^{-29}$    &   5.8$\times$10$^{-9}$    &     \\
 CeIrIn$_5$	        & 0.4    &  1.27$\times$10$^{-28}$    &   2.41$\times$10$^{-8}$    &     \\ 
 PrOs$_4$Sb$_{12}$         & 1.8    &  9.1$\times$10$^{-29}$    &   1.2$\times$10$^{-8}$    &          \\ 
\hline
 YBa$_2$Cu$_3$O$_{7-\delta}$	 & 51.5 	&  3.91$\times$10$^{-30}$    &   2.82$\times$10$^{-9}$   &  \cite{ando02,ybco-mass,ramshaw15,legros18}     \\
 				             & 54.1 	&  2.64$\times$10$^{-30}$    &   2.92$\times$10$^{-9}$   &       \\
 				           & 54.7 	&  2$\times$10$^{-30}$    &   2.98$\times$10$^{-9}$   &       \\
				           & 56.2  	&  1.27$\times$10$^{-30}$    &   3.1$\times$10$^{-9}$   &       \\
 				           & 59.2 	&  1.02$\times$10$^{-30}$    &   3.33$\times$10$^{-9}$   &       \\
 				           & 61.4 	&  8.55$\times$10$^{-31}$    &   3.72$\times$10$^{-9}$   &       \\
				           & 65.9 	&  1.27$\times$10$^{-30}$    &   3.7$\times$10$^{-9}$   &       \\
				           & 77.7  	&  1.91$\times$10$^{-30}$    &   3.07$\times$10$^{-9}$   &       \\
				           & 84.7 	&  2.18$\times$10$^{-30}$    &   2.7$\times$10$^{-9}$   &       \\
				           & 92	 	&  3.27$\times$10$^{-30}$    &   2.17$\times$10$^{-9}$   &       \\
 Bi$_2$Sr$_2$CaCu$_2$O$_{8+\delta}$	           & 50  	&  2.7$\times$10$^{-30}$    &   1.51$\times$10$^{-9}$   &  \cite{ong03,bisco-mass}     \\
 									           & 77  	&  2.4$\times$10$^{-30}$    &   1.81$\times$10$^{-9}$   &       \\
											   & 90  	&  2.3$\times$10$^{-30}$    &   2.22$\times$10$^{-9}$   &       \\
									           & 78  	&  2.2$\times$10$^{-30}$    &   2.31$\times$10$^{-9}$   &       \\
									           & 65  	&  2$\times$10$^{-30}$    &   2.56$\times$10$^{-9}$   &       \\
 La$_{2-x}$Sr$_x$CuO$_4$ 	             & 40 	&  3.6$\times$10$^{-30}$    &   2.5$\times$10$^{-9}$    &  \cite{cuprates}     \\
 Tl$_2$Ba$_2$CaCu$_2$O$_{8+x}$ 	         & 97 	&  7.6$\times$10$^{-30}$    &   3$\times$10$^{-9}$    &  \cite{cuprates}     \\
 HgBa$_2$Ca$_2$Cu$_3$O$_{8+x}$ 	         & 135 	&  2.7$\times$10$^{-30}$    &   1.5$\times$10$^{-9}$    &  \cite{ong03}     \\
 Nd$_{2-x}$Ce$_x$CuO$_4$ 	             & 24.5 	&  3.6$\times$10$^{-31}$    &   7.5$\times$10$^{-9}$    &  \cite{ong03}     \\
 \hline
 NdFeAsO$_{0.7}$F$_{0.3}$    		    & 47    &  1.82$\times$10$^{-30}$     &   2.47$\times$10$^{-9}$    &  \cite{johnston10}        \\ 
 Ba$_{0.6}$K$_{0.4}$Fe$_2$As$_2$         & 28.2  &  4.34$\times$10$^{-30}$     &   2.4$\times$10$^{-9}$    &          \\ 
 Ba(Fe$_{0.9}$Co$_{0.1}$)$_2$As$_2$      & 22    &  4.55$\times$10$^{-30}$     &   3.63$\times$10$^{-9}$    &          \\ 
 Sr(Fe$_{0.9}$Co$_{0.1}$)$_2$As$_2$      & 20    &  4.55$\times$10$^{-30}$     &   2.67$\times$10$^{-9}$    &          \\ 
 FeSe$_{0.6}$Te$_{0.4}$       		    & 14    &  2.73$\times$10$^{-30}$     &   2.67$\times$10$^{-9}$    &          \\ 
 \hline
$\beta$--(ET)$_2$I$_3$	       & 1.1 	&  4.23$\times$10$^{-30}$    &   6.33$\times$10$^{-8}$   &  \cite{organicsc}     \\
$\beta$--(ET)$_2$IBr$_2$  	   & 2.25 	&  3.82$\times$10$^{-30}$    &   4.63$\times$10$^{-8}$   &      \\
$\beta$--(ET)$_2$AuI$_2$        & 4.2  	&  32$\times$10$^{-31}$      &   2.49$\times$10$^{-8}$   &       \\
(TMTSF)$_2$ClO$_4$       		& 1.25 	&  32$\times$10$^{-31}$      &   7.06$\times$10$^{-8}$   &  		     \\
$\kappa$--(ET)$_2$Cu(NCS)$_2$ 	& 8.3 	&  3.18$\times$10$^{-30}$    &   7$\times$10$^{-9}$      &       \\
$\kappa$--(ET)$_2$[N(CN)$_2$]Br  & 11.3 	&  8.65$\times$10$^{-31}$    &   3.7$\times$10$^{-9}$    &       \\
\hline
 H$_2$S	           & 203    &     2.28$\times$10$^{-31}$  &   2.15$\times$10$^{-9}$   &  \cite{drozdov15}   \\ 
\hline
 CNT               & 15    &     3.3$\times$10$^{-31}$   &   4.2$\times$10$^{-9}$  &  \cite{cnt}     \\   
\hline
 MATBG             & 1.7    &     9.1$\times$10$^{-32}$   &   5.2$\times$10$^{-8}$  &  \cite{matbg}     \\ 
 MATTG	           & 1.2    &     9.1$\times$10$^{-31}$   &   3.8$\times$10$^{-8}$   &  \cite{mattg}   \\ 
 WSe2              & 2$\times$10$^{-1}$   &     9.1$\times$10$^{-30}$   &   5.2$\times$10$^{-8}$   &  \cite{wse2}   \\ 
\hline
 Cu$_x$Bi$_2$Se$_3$	& 3.8    &     2.4$\times$10$^{-30}$  &   1.39$\times$10$^{-8}$   &  \cite{topologicalsc}   \\  
\hline
 Bi             & 5.3$\times$10$^{-4}$     &     9.1$\times$10$^{-34}$   &   9.6$\times$10$^{-5}$  &  \cite{bismuth}     \\ 
 YbRh$_2$Si$_2$ & 7.9$\times$10$^{-3}$     &     1$\times$10$^{-27}$     &   9.7$\times$10$^{-8}$   &  \cite{ybrh2si2}   \\ 
\hline
 K$_2$Cr$_3$As$_3$   & 6.2     &     1.6$\times$10$^{-30}$     &   3.5$\times$10$^{-9}$   &  \cite{1dsc}   \\ 
\hline
 Li$_x$ZrNCl	       & 15.9 	&  8.19$\times$10$^{-31}$    &   5.41$\times$10$^{-9}$   &  \cite{ZrNCl}     \\
 			       & 16.1 	&  8.19$\times$10$^{-31}$    &   5.56$\times$10$^{-9}$   &       \\
  		           & 17.8 	&  8.19$\times$10$^{-31}$    &   6.80$\times$10$^{-9}$   &      \\
 		           & 18.1  	&  8.19$\times$10$^{-31}$    &   7.71$\times$10$^{-9}$   &       \\
 		           & 19.0 	&  8.19$\times$10$^{-31}$    &   6.80$\times$10$^{-9}$      &       \\
 		           & 15.9 	&  8.19$\times$10$^{-31}$    &   7.42$\times$10$^{-9}$    &       \\
 		           & 13.1 	&  8.19$\times$10$^{-31}$    &   1.23$\times$10$^{-8}$    &       \\
 		           & 11.5 	&  8.19$\times$10$^{-31}$    &   2.1$\times$10$^{-8}$   &       \\


\end{longtable}


\begin{table}[b]
\caption{The critical temperature T$_c$, normal state effective mass 
$m_{eff}$ and coherence length $\xi$ for various superfluids and condensates.}
\begin{tabular}{|l|c|c|c|l|l|}
\hline
 Condensate  & $T_c$ ($K$)  & $m_{eff}$ ($kg$)   & $\xi$ ($m$)  & Reference  & Type \\ \hline
 $^4$He         & 2.17    &  1.135$\times$10$^{-26}$    &   4$\times$10$^{-10}$   &  \cite{he-4-mass,he-4-length}   & bosonic     \\ 
\hline
 $^3$He         &    					   &  					       &   					      &  \cite{he3-pressure}  & fermionic       \\ 
  p = 0 bar     &  0.929$\times$10$^{-3}$   &  1.40$\times$10$^{-26}$    &   7.72$\times$10$^{-8}$   &  			 &		        \\ 
  p = 2 bar     &  1.181$\times$10$^{-3}$   &  1.53$\times$10$^{-26}$    &   5.70$\times$10$^{-8}$   &              &              \\ 
  p = 4 bar     &  1.388$\times$10$^{-3}$   &  1.64$\times$10$^{-26}$    &   4.59$\times$10$^{-8}$   &              &              \\ 
  p = 6 bar     &  1.560$\times$10$^{-3}$   &  1.74$\times$10$^{-26}$    &   3.88$\times$10$^{-8}$   &              &              \\ 
  p = 8 bar     &  1.705$\times$10$^{-3}$   &  1.84$\times$10$^{-26}$    &   3.40$\times$10$^{-8}$   &              &              \\ 
  p = 10 bar     &  1.828$\times$10$^{-3}$   &  1.93$\times$10$^{-26}$    &   3.04$\times$10$^{-8}$   &             &               \\ 
  p = 12 bar     &  1.934$\times$10$^{-3}$   &  2.02$\times$10$^{-26}$    &   2.77$\times$10$^{-8}$   &             &               \\ 
  p = 14 bar     &  2.026$\times$10$^{-3}$   &  2.10$\times$10$^{-26}$    &   2.55$\times$10$^{-8}$   &             &               \\ 
  p = 16 bar     &  2.106$\times$10$^{-3}$   &  2.19$\times$10$^{-26}$    &   2.38$\times$10$^{-8}$   &             &               \\ 
  p = 18 bar     &  2.177$\times$10$^{-3}$   &  2.27$\times$10$^{-26}$    &   2.23$\times$10$^{-8}$   &             &               \\ 
  p = 20 bar     &  2.239$\times$10$^{-3}$   &  2.35$\times$10$^{-26}$    &   2.10$\times$10$^{-8}$   &             &               \\ 
  p = 22 bar     &  2.293$\times$10$^{-3}$   &  2.43$\times$10$^{-26}$    &   1.99$\times$10$^{-8}$   &             &               \\ 
  p = 24 bar     &  2.339$\times$10$^{-3}$   &  2.51$\times$10$^{-26}$    &   1.90$\times$10$^{-8}$   &             &               \\ 
  p = 26 bar     &  2.378$\times$10$^{-3}$   &  2.59$\times$10$^{-26}$    &   1.82$\times$10$^{-8}$   &             &               \\ 
  p = 28 bar     &  2.411$\times$10$^{-3}$   &  2.67$\times$10$^{-26}$    &   1.74$\times$10$^{-8}$   &             &               \\ 
  p = 30 bar     &  2.438$\times$10$^{-3}$   &  2.75$\times$10$^{-26}$    &   1.68$\times$10$^{-8}$   &             &               \\ 
  p = 32 bar     &  2.463$\times$10$^{-3}$   &  2.83$\times$10$^{-26}$    &   1.62$\times$10$^{-8}$   &             &               \\ 
  p = 34 bar     &  2.486$\times$10$^{-3}$   &  2.91$\times$10$^{-26}$    &   1.58$\times$10$^{-8}$   &             &               \\ 
\hline
 $^1$H              & 5$\times$10$^{-5}$    &  1.66$\times$10$^{-27}$    &   1.85$\times$10$^{-6}$   &  \cite{kleppner98}   & bosonic      \\ 
 $^7$Li             & 3$\times$10$^{-7}$   &  1.16$\times$10$^{-26}$    &   4.3$\times$10$^{-6}$   &  \cite{kleppner98}    &    \\ 
 $^{23}$Na	        & 2$\times$10$^{-6}$ 	&  3.82$\times$10$^{-26}$    &   3.1$\times$10$^{-7}$    &  \cite{kleppner98}   &    \\
 $^{87}$Rb	        & 6.7$\times$10$^{-7}$  &  1.44$\times$10$^{-25}$    &   1.88$\times$10$^{-7}$     &  \cite{kleppner98} &   \\ 
\hline
 $^6$Li	           & 5$\times$10$^{-8}$ 	&  9.98$\times$10$^{-27}$    &   1.8$\times$10$^{-5}$   &  \cite{bec-bcs}   & fermionic  \\
 \hline
exciton            & 1     &  1$\times$10$^{-31}$    &   1$\times$10$^{-6}$    &  \cite{High2012}   & bosonic   \\ 
polariton      & 300   &  1$\times$10$^{-35}$    &   1$\times$10$^{-6}$     &  \cite{polariton-mass,polariton-length,polariton-Tc}  & bosonic   \\ 
\hline 
neutron superfluid    & 1.16$\times$10$^{10}$   &  1.673$\times$10$^{-27}$     &   1$\times$10$^{-14}$    &  \cite{bcs50}  & fermionic \\ 
proton superconductor  & 4.64$\times$10$^{9}$   &  1.675$\times$10$^{-27}$     &   1$\times$10$^{-14}$    &  \cite{bcs50}  & fermionic \\ 
\hline
nuclear pairing          & 5.8$\times$10$^{9}$   &  1.67$\times$10$^{-27}$      &   2.7$\times$10$^{-14}$    &  \cite{nuclear_sf} & fermionic  \\ 
\hline
 quark-antiquark condensate & 1.74$\times$10$^{12}$ & 2.4$\times$10$^{-28}$  &   1.61$\times$10$^{-15}$ & \cite{quark-gluon-book} & fermionic    \\ 
\hline
 Higgs condensate     & 1.85$\times$10$^{15}$   &  2.23$\times$10$^{-25}$    &   1.5$\times$10$^{-18}$    &  \cite{coleman-book}  & bosonic   \\ 
\hline
 Planck              & 1.41$\times$10$^{32}$   &  2.17$\times$10$^{-8}$    &   1.61$\times$10$^{-35}$    &  \cite{planck-book}   &   \\ 
\hline
\end{tabular}
\label{tab:sfdatabase}
\end{table}

%


\clearpage


\begin{figure*}[t]
\vspace*{-1.0cm}%
\centerline{\includegraphics[width=9in,angle=90]{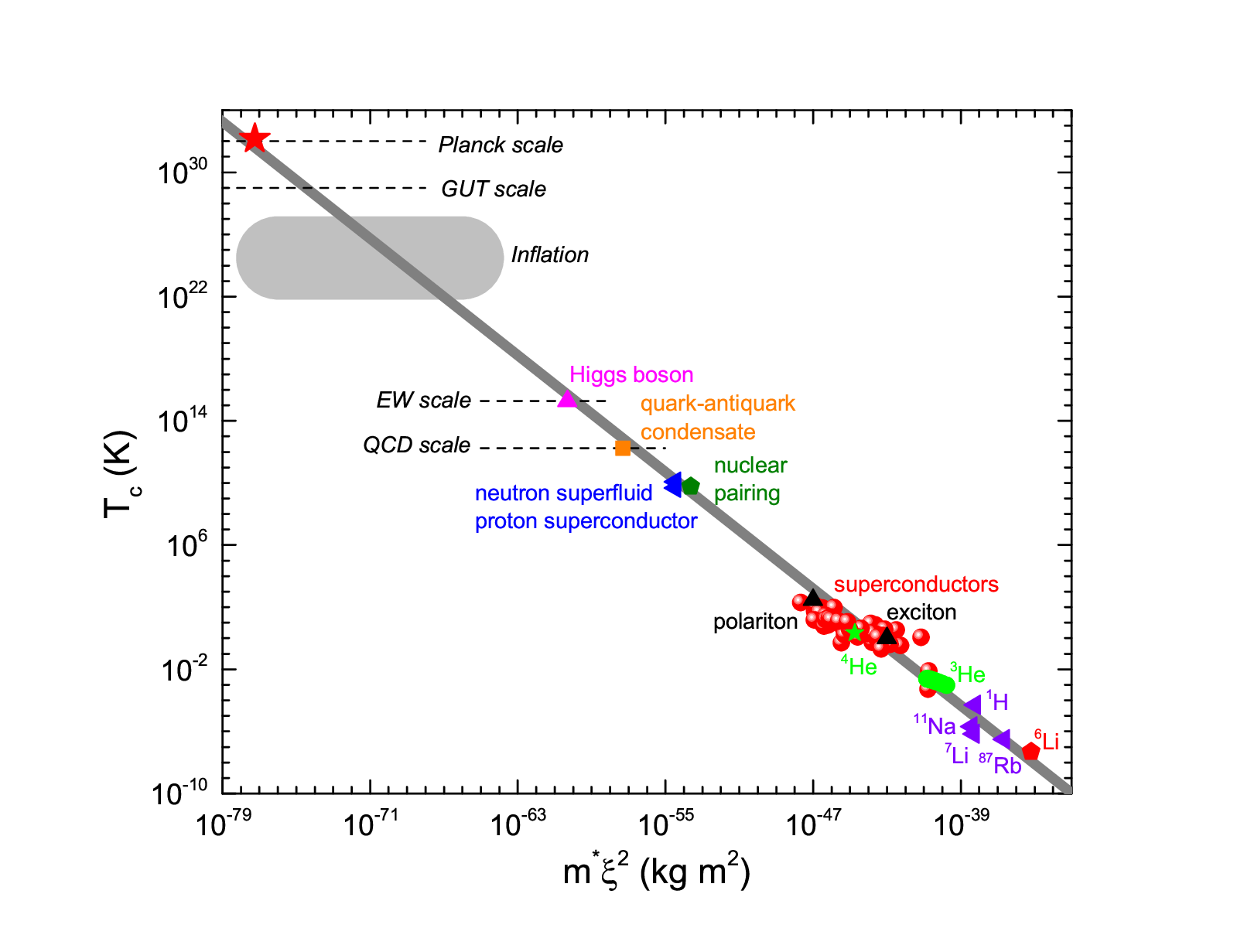}}%
\vspace*{-2.0cm}%
\caption{The critical temperature T$_c$ (in $K$) versus 
$m^{\ast}\xi^2$, where $m^{\ast}$ (in $kg$) is the mass of condensing particles and
$\xi$ (in $m$) is the zero temperature coherence length for various quantum systems 
undergoing condensation (Tables \ref{tab:scdatabase} and \ref{tab:sfdatabase}). 
Red circles represent superconductors (Table \ref{tab:scdatabase}). 
Several characteristic temperature (energy) scales are also shown: Planck, 
GUT, inflation, EW and QCD.}
\vspace*{0.0cm}%
\label{fig:all}
\end{figure*}


\end{document}